\documentclass{kapproc} 

\usepackage{procps} 

\usepackage[dvips]{graphicx}

\upperandlowercase

\kluwerbib 

\begin{document}

\articletitle[The End of the Dark Ages]
{The End of the Dark Ages:\\
Probing the Reionization of the Universe \\
with HST and JWST}

\author{N. Panagia [ESA/STScI], M. Stiavelli \& S.M. Fall [STScI]}

\email{panagia@stsci.edu; mstiavel@stsci.edu; fall@stsci.edu}

\chaptitlerunninghead{The Reionization of the Universe}

\anxx{Panagia\, Nino}
\anxx{Stiavelli\, Massimo}
\anxx{Fall\, S. Michael}

\begin{abstract}
Limiting the number of model-dependent  assumptions to a minimum, we
discuss the detectability of the sources  responsible for reionization
with existing and planned telescopes. We conclude that if reionization
sources are UV-efficient, minimum luminosity sources, then it may be 
difficult to detect them before the advent of the James Webb Space
Telescope (JWST).  The best approach before the launch of JWST may be
either to  exploit gravitational lensing by clusters of galaxies, or to
search for strong Ly-$\alpha$ sources by means of narrow-band excess
techniques or slitless grism spectroscopy.

\end{abstract}


\section{Introduction}

Motivated by recent evidence that the epoch of reionization of hydrogen
may have ended at a redshift as low as $z\approx 6$ ({\it
e.g.,~}\cite{becker01,fan02}), we have considered the detectability of 
the sources responsible for this reionization. The main idea is that
reionization places limits on the mean surface brightness of the
Population of reionization sources. We have defined a family of models
characterized by two parameters: the Lyman continuum escape fraction
$f_c$ from the sources, and the clumpiness parameter $C$ of the
intergalactic medium.  The minimum surface brightness model corresponds
to a value of unity for both parameters. A maximum surface brightness
is obtained by requiring that the reionization sources do not
overproduce heavy elements. Our general approach is applicable to most
types of reionization sources, but in specific numerical examples we
focus on  Population III stars, because they have very high effective
temperatures and, therefore, are very effective producers of ionizing
UV photons ({\it e.g.,~} \cite{panagia03}). Our mean surface brightness
estimates are compared to the parameter space that can be probed by
existing and future telescopes, in order to help  planning the most
effective surveys.  A full account of our work can be found in
\cite{stiavelli03}.

\begin{figure} \centering
\includegraphics[height=7.cm]{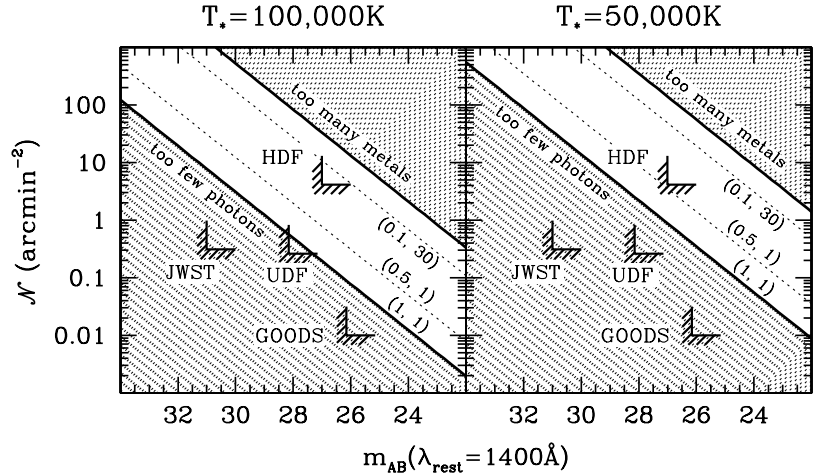}
\caption{The loci of surface density vs apparent AB magnitude for
identical reionization sources that are either Population III (left hand
panel) or Population II stars (right hand panel). 
\label{Fig.1}}  
\end{figure} 

\section{Results and Discussion}

\begin{figure} 
\centering 
\includegraphics[height=8.4cm]{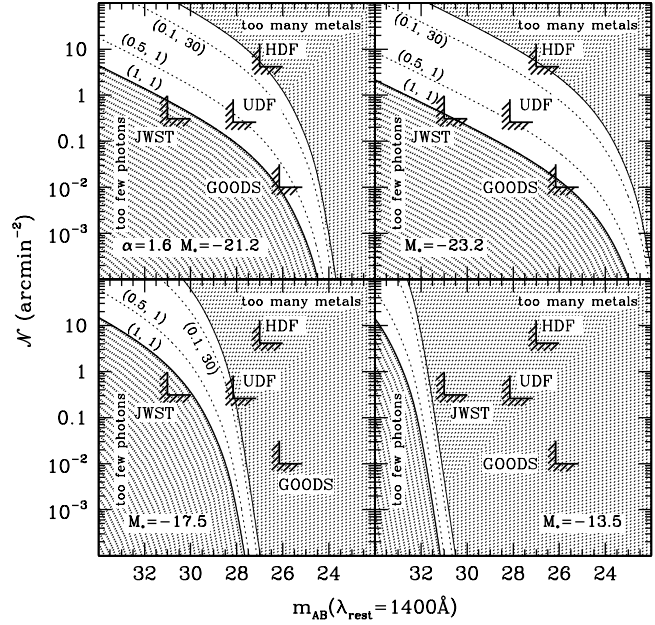}
\caption{The cumulative distribution of the surface density vs apparent
AB magnitude of reionization sources with luminosity functions with
different knees.   
\label{fig2}} 
\end{figure}

In Figure 1 we show the loci of the mean surface brightness of {\it
identical} reionization sources  as a function of their observable AB
magnitude. The left panel refers to Population III sources with effective
temperature of $10^5$ K,  the right panel to Population II reionization
sources with effective temperature $5 \times 10^4$ K. In both panels,
the  lower solid line represents the minimum surface brightness model,
(1,1), while the upper solid line represents the global metallicity
constraint $Z\leq 0.01 Z_\odot$ at $z=6$. The thin dotted lines
represent the (0.5, 1) and (0.1, 30) models.  The non-shaded area is
the only one accessible to reionization sources that do not overproduce
metals. The L-shaped markers delimit the quadrants ({i.e.,} the areas
above and to the right of the markers) probed by the GOODS/ACS survey
(\cite{Giavalisco}), the HDF/HDFS NICMOS fields
(\cite{Thompson,WilliamsHDFS}), the Ultra Deep Field (UDF)  and by a
hypothetical ultra-deep survey with JWST.

In Figure 2 we show the expected cumulative surface density
distributions for reionization sources with a variety of   luminosity
functions. In each panel, the upper solid line represents the global
metallicity constraint. The lower solid line represents the minimum
surface brightness model, (1,1). The thin dotted lines give the
luminosity function for the (0.5,1) and the (0.1, 30) models. The
L-shaped markers delimit the quadrants probed by the GOODS/ACS survey,
the HDF and HDFS NICMOS fields, the UDF, and an ultra-deep survey with
JWST, respectively.

From these results it appears that if reionization is caused by
UV-efficient, minimum surface brightness sources, the non-ionizing
continuum emission from reionization sources will be difficult to
detect before the advent of JWST.  On the other hand, if the sources of
reionization were not extremely hot Population III stars but cooler
Population II stars or AGNs, they would be brighter by 1-2 magnitudes
and thus they would be easier to detect. 

\begin{figure} 
\centering 
\includegraphics[height=7.5cm]{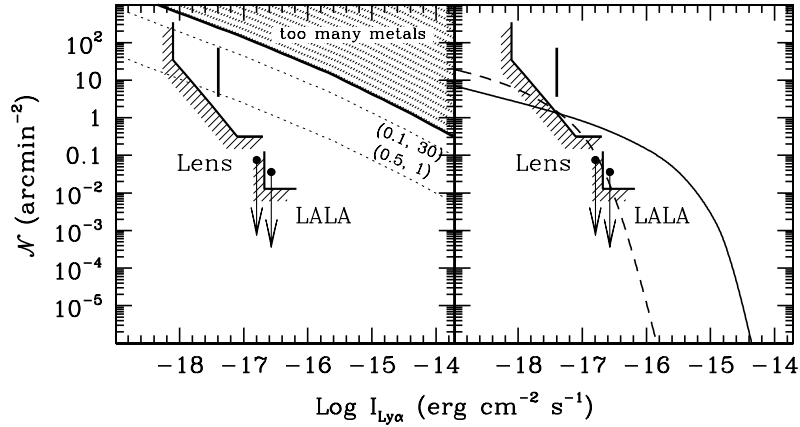}
\caption{Surface density vs Ly-$\alpha$ line flux of reionization
sources.   
\label{fig3}} 
\end{figure}

Finally, Figure 3 presents the required surface density as a function
of the Ly-$\alpha$ line flux of reionization sources. The left hand
panel shows the loci of  identical sources  for two different $(f_c,
C)$ models (thin dotted lines).  The top solid line identifies the
global metallicity constraint.  The right hand panel illustrates two
different luminosity functions. The solid line refer to a luminosity
function identical to that of $z=3$ Lyman break galaxies, ({\it i.e.},
$M_{*,1400}=-21.2$ and $\alpha=1.6$,  while the dashed line refers to a
luminosity function with $M_{*,1400}=-17.5$ and a slope identical to
the local Universe slope, $\alpha=1.1$. In both panels, the L-shaped
marker identifies the quadrant probed by a narrow-band excess survey at
$z \geq 6$ (LALA survey, \cite{rhoads01}). The oblique marker labeled
{\it Lens} represents a hypothetical 100-orbit survey with the ACS
grism on a cluster of galaxies to exploit gravitational amplification.
The solid bar represents the density estimated from the detection at
$z=6.56$ by \cite{hu02} while the two points with down-pointing arrows
represent their upper limits. 

It appears that searches based on narrow-band excess techniques or
slitless grisms would be promising and might lead to the detection of
the reionization sources within this decade.


\begin{chapthebibliography}{1}

\bibitem[Becker et al. (2001)]{becker01} Becker, R.H., et al. 2001, AJ,
122, 2850

\bibitem[Fan et al. (2002)]{fan02} Fan, X., et al. 2002, AJ, 123, 1247

\bibitem[Dickinson and Giavalisco (2003)]{Giavalisco} Dickinson, M., \&
Giavalisco, M. 2003, in the proceedings of the ESO/USM Workshop "The
Mass of Galaxies at Low and High Redshift", eds.  R. Bender \& A.
Renzini, p. 324

\bibitem[Hu et al. (2002)]{hu02} Hu, E.M., et al., 2002, ApJ, 568, L75

\bibitem[Panagia et al. (2003)]{panagia03} Panagia, N., Stiavelli, M.,
Ferguson, H. \& Stockman, H.S., 2003, in preparation [see also 
astro-ph/0209278 and /0209346]

\bibitem[Rhoads and Malhotra (2001)]{rhoads01} Rhoads, J.E. \&
Malhotra, S., 2001, ApJ, 563, L5

\bibitem[Stiavelli, Fall \& Panagia (2003)]{stiavelli03} Stiavelli, M.,
Fall, S.M., \& Panagia, N., 2003, ApJ, in press

\bibitem[Thompson et al. (1999)]{Thompson} Thompson, R.,
Storrie-Lombardi, L., Weymann, R.J.,  Rieke, M.J., Schneider, G.,
Stobie, E., \& Lytle, D., 1999, AJ, 117, 17

\bibitem[Williams et al. (2000)]{WilliamsHDFS} Williams, R.E., et al.,
2000, AJ, 120, 2735

\end{chapthebibliography}
\end{document}